\documentclass[a4paper,fleqn]{cas-dc}

\usepackage[numbers,sort&compress]{natbib}

\usepackage{float}\restylefloat{figure}
\usepackage{placeins}  
\usepackage{soul}
\usepackage{upgreek}
\renewcommand{\pi}{\uppi}

\usepackage{amsmath} 

\def\tsc#1{\csdef{#1}{\textsc{\lowercase{#1}}\xspace}}
\tsc{WGM}
\tsc{QE}

\begin{document}
\let\WriteBookmarks\relax

\title [mode = title]{Non-equilibrium evaporation of Lennard-Jones fluids: Enskog-Vlasov Equation and Hertz-Knudsen model}  

\author[1,2]{Shaokang Li}

\affiliation[1]{organization={Center for Interdisciplinary Research in Fluids, Institute of Mechanics, Chinese Academy of Sciences},
            city={Beijing},
            postcode={100190}, 
            country={PR China}}

\affiliation[2]{organization={School of Engineering, The University of Edinburgh},
            city={Edinburgh},
            postcode={EH9 3FQ}, 
            country={UK}}

\author[2]{Livio Gibelli}

\author[1,3]{Yonghao Zhang}


\affiliation[3]{organization={School of Engineering Science, University of Chinese Academy of Sciences}, 
            city={Beijing},
            postcode={101408}, 
            country={PR China}}

\cortext[2]{Corresponding author}

\ead{yonghao.zhang@imech.ac.cn}

\begin{abstract}
Enskog-Vlasov equation is currently the most sophisticated kinetic model for describing non-equilibrium evaporative flows. While it enables more efficient simulations than the molecular dynamics (MD) methods, its accuracy in reproducing the flow properties of real fluids is limited by both the assumptions underlying the Vlasov forcing term and the approximation introduced by the Enskog collision term for short-range molecular interactions. To address this limitation, this work proposes a molecular kinetic model specifically designed for real fluids, with the Lennard-Jones fluids as an example. The model is first applied to evaluate the equilibrium characteristics of a liquid-vapour system, including the liquid-vapour coexistence curve, transport coefficients, vapour pressure, and surface tension coefficient. The results show excellent agreement with the MD simulation and experimental data. Furthermore, the model is used to investigate non-equilibrium evaporation, with a particular focus on the velocity distribution function adjacent to the liquid-vapour interface. The results confirm that deviations from the Maxwellian distribution persist in the vapour region, indicating limitations of the classical Hertz–Knudsen relation under pronounced non-equilibrium conditions. This work represents a critical step towards the development of an accurate and efficient computational framework for modelling non-equilibrium liquid-vapour flows for real fluids, with direct relevance to practical applications such as flow cooling.

\end{abstract}

\begin{highlights}
\item A simplified kinetic model is developed to accurately simulate non-equilibrium evaporative flow of Lennard-Jones fluids.

\item The model successfully reproduces key equilibrium properties, including the liquid-vapour coexistence curve,  transport coefficient, vapour pressure, and surface tension.

\item The velocity distribution function reveals significant non-Maxwellian features adjacent to the liquid-vapour interface, suggesting the breakdown of the classical Hertz–Knudsen relation under strong evaporation conditions.
\end{highlights}



\maketitle

\section{Introduction}

Nanoscale evaporation plays a crucial role in advancing nanotechnologies, ranging from improving substance separation efficiency in nanoporous membranes to enabling highly effective heat dissipation in high performance electronic devices \citep{vaartstra2020capillary, dong2022high, wang2025molecular}. However, accurate modelling of nano-scale evaporation presents a significant research challenge due to the involvement of widely separated physical length and time scales. 

One of the earliest and most widely-used approaches to describe the evaporation and condensation processes is the Hertz-Knudsen (HK) relation. Derived from kinetic theory, this empirical relation assumes local equilibrium in the vapour region and describes the phase change flux (evaporation or condensation) at the interface between a liquid and its vapour, serving as a cornerstone in classical models of interfacial mass transfer \citep{hertz1882ueber, knudsen1915maximale}. Its simplicity makes it particularly suited for practical applications, where detailed molecular transport effects are not the primary focus. However, at the nano/micro-scale, the HK relation overlooks the kinetic structure of the evaporation process in the Knudsen layer, a region adjacent to the liquid-vapour interface where non-equilibrium effects, such as velocity slip and temperature jump, arise due to molecular interactions \citep{frezzotti2007numerical}. The Schrage equation extends the classical HK formula by incorporating non-equilibrium effects. However, Hertz-Knudsen-Schrage framework is valid for slow flows (low Mach numbers), and the Knudsen layer is ignored \citep{schrage1953theoretical, ytrehus1996kinetic, vaartstra2022revisiting, oskouei2025nonlinear}.

To accurately capture the complex non-equilibrium effects within the Knudsen layer, the kinetic approach is required, as it directly accounts for molecular collisions and the kinetic structure of the evaporation process. For evaporation from a planar liquid surface, the structure of the Knudsen layer and the jump relationship along this region have been well described theoretically using the moment method to solve the Boltzmann model equations \citep{labuntsov1979analysis, ytrehus1997molecular, meland2003evaporation}. In parallel with these theoretical investigations, numerical simulations based on the Boltzmann equation or related kinetic models have been developed to tackle this challenge. For instance, the structure of the Knudsen layer during evaporation from a planar surface was examined using the Shakhov model \citep{graur2021non}. Meanwhile, the density and temperature drop in two-dimensional geometries, such as nanoporous membranes, were investigated through direct simulation Monte Carlo (DSMC) method \citep{john2019numerical, john2021evaporation, li2021theoretical, li2023effect}. Although these approaches are well established, they typically model only the vapour phase explicitly and do not fully resolve the structure and dynamics of the liquid–vapour interface. As a result, the molecular exchange process with the liquid phase depends on a phenomenological boundary condition that requires an evaporation coefficient as an input. However, reported values for this coefficient vary by up to three orders of magnitude, introducing significant uncertainty \citep{persad2016expressions}.

To overcome the reliance on the arbitrary evaporation coefficient, one promising approach is to develop a unified framework that cohesively describes the liquid phase, vapour phase, and their interface. The key challenge lies in accounting for the volume exclusion and molecular attraction between fluid molecules. The former has been addressed by the Enskog theory, which extends the Boltzmann equation to dense fluids, modifying the collision frequency and incorporating non-localized binary collisions. The latter is modelled by introducing a mean-field Vlasov forcing term into the Enskog equation to account for long-range forces between fluid molecules. The resulting Enskog–Vlasov (EV) equation provides a unified framework for describing non-equilibrium flows in both the liquid and vapour phases, as well as the evaporation process at the liquid–vapour interface. A notable advantage of the EV equation is that it eliminates the need for traditional phase change models, which rely on evaporation or condensation coefficients \citep{enskog1922kinetische, sobrino1967kinetic, grmela1971kinetic, karkheck1981kinetic}. Recent studies have further demonstrated that the EV equation can describe liquid–vapour flows across a wide range of fluids and flow conditions \citep{frezzotti2005mean, barbante2015kinetic, frezzotti2017kinetic, busuioc2020mean}. 

Although the EV equation can qualitatively capture liquid–vapour flow behavior, its underlying assumptions limit its accuracy in reproducing the flow properties of real fluids. In this work, we take the Lennard-Jones fluid, defined as particles interacting via the standard 12-6 Lennard-Jones potential, as an example to demonstrate how real fluids can be considered in the Enskog-Vlasov model. The 12-6 Lennard-Jones potential is expressed as \citep{tee1966molecular, hansen1969phase, meland2003evaporation, chen2024evaporation}:
\begin{equation}
   \phi(r) = \phi_{{\rm LJ}}\left[\left(\frac{\sigma}{r}\right)^{12}-\left(\frac{\sigma}{r}\right)^{6}\right],
\label{eq 1}
\end{equation}
where $\phi_{{\rm LJ}}$ represents the potential well, $\sigma$ is the molecular diameter, and $r$ denotes the molecular distance. By contrast, in the EV equation, molecular interactions are modelled using the Sutherland potential. This assumes infinitely strong repulsion when the molecular distance is less than one diameter, reflecting the hard sphere approximation in the Enskog theory, while the attractive interactions are represented by a mean-field approximation of the attractive tail of the Lennard-Jones potential, i.e.,
\begin{align}
   \phi(r) = \left\{
   \begin{aligned}
   &+\infty, \quad  &r&< \sigma, \\
   & -\phi_{\sigma} \left(\frac{\sigma}{r}\right)^{6}, &\quad  r &\geq \sigma.
   \end{aligned}
   \right.
\label{eq 2}
\end{align}
where $\phi_{\sigma}$ represents an adjusted constant parameter. It is evident that the Lennard-Jones and Sutherland potentials exhibit markedly different repulsive interaction characteristics, leading to notable difference in the fluid behaviour. For example, the critical number density $n_\mathrm{c}$ and critical temperature $T_{\mathrm{c}}$ of Lennard-Jones fluids are \citep{chung1988generalized}:
\begin{equation}
    n_{\mathrm{c}}\sigma^3  \approx 0.3185, \quad T_{\mathrm{c}} = 0.3148\frac{\phi_{{\rm LJ}}}{k_{\rm B}},
\label{eq 3}
\end{equation}
where $k_{\rm B}$ is the Boltzmann constant, whereas these parameters are calculated through
\begin{equation}
    n_\mathrm{c}\sigma^3  \approx 0.2484, \quad T_\mathrm{c} = 0.752\frac{\phi_{\sigma}}{k_{\rm B}}
\label{eq 4}
\end{equation}
in the EV framework. Although the critical temperature can be matched by adjusting $\phi_{\sigma}$, the critical reduced number density differs substantially. Therefore, it is necessary to make appropriate modifications to the EV equation to ensure consistency with the thermodynamic properties of Lennard-Jones fluids, i.e., the critical and triple-point densities and temperatures. 

Although many efforts have been made to resolve this issue, there is still no kinetic model capable of accurately reproducing the two-phase flow properties of Lennard-Jones fluids with an acceptable computational cost \citep{benilov2018energy, benilov2019peculiar,homes2025heat}. Based on the simplified kinetic model developed in our previous studies, we propose a modification to the pair correlation function to ensure the correct equation of state for Lennard-Jones fluids \citep{wang2020kinetic, su2023kinetic, shan2023molecular, li2024molecular, shan2025molecular}. The resulting model can then reproduce critical properties of the Lennard-Jones fluids. The proposed model will first be assessed for its accuracy, including the liquid–vapour coexistence curve, vapour pressure, and surface tension coefficient. The results will also be compared with MD and experimental data to confirm whether the model can reliably predict the properties of Lennard-Jones fluids. The model will finally be employed to investigate non-equilibrium evaporation, providing detailed insights into the velocity distribution function at the interface and in the Knudsen layer. We will show that the deviations from the Maxwellian distribution persist in the vapour region, indicating limitations of the classical Hertz–Knudsen relation under pronounced non-equilibrium conditions.

The remainder of the letter is organised as follows. Section \ref{section2} presents the simplified kinetic model and the modifications introduced to reproduce the equation of state of a Lennard-Jones fluid. Section \ref{section 3} describes the simulations conducted under both equilibrium and non-equilibrium conditions. Finally, Section \ref{section4} summarizes the main findings and discusses future research directions.

\section{Molecular Kinetic Model}
\label{section2}

We consider a fluid composed of identical atoms interacting via the Sutherland potential (Eq. (\ref{eq 2})). The statistical behavior of this fluid is described by the molecular velocity distribution function $f(\boldsymbol{x}, \boldsymbol{\xi}, t)$, which represents the number of atoms at time $t$ within an infinitesimal volume of single-particle phase space centered at position $\boldsymbol{x}$ and molecular velocity $\boldsymbol{\xi}$. Following the Enskog theory for dense gases and the mean-field theory for molecular interaction force, a closed-form evolution equation for the distribution function can be derived, commonly known as the EV equation:
\begin{equation}
   \frac{\partial f}{\partial t} + \boldsymbol{\xi} \cdot \frac{\partial f}{\partial \boldsymbol{x}} +\frac{\boldsymbol{F}}{m} \cdot \frac{\partial f}{\partial \boldsymbol{\xi}} = \varOmega, 
\label{eq 5}
\end{equation}
where $\boldsymbol{F}$ is the self-consistent force field generated by the Sutherland potential:
\begin{equation}
   \boldsymbol{F}(\boldsymbol{x},t) = \int_{||\boldsymbol{x_{1}}-\boldsymbol{x}|| > \sigma}\frac{\mathrm{d}\phi}{\mathrm{d}r}\frac{\boldsymbol{x}_{1}-\boldsymbol{x}}{||\boldsymbol{x}_{1}-\boldsymbol{x}||}n(\boldsymbol{x}_{1},t)\mathrm{d}\boldsymbol{x}_{1},
\label{eq 6}
\end{equation}
and $\varOmega$ denotes the hard-sphere collision term derived from the Enskog theory.

While the Enskog collision term can accurately capture the short-range interactions of dense gases, its high computational cost limits its applicability in practical engineering simulations. To address this challenge, we introduce a simplified kinetic model by approximating the non-local Enskog collision term using a first-order Taylor expansion in molecular diameter. The zeroth-order term is replaced by the Shakhov model to ensure the correct Prandtl number (Pr), and the first-order derivatives are evaluated by substituting the distribution function with a Maxwellian. Additionally, a second-order velocity term is retained to recover the correct bulk viscosity \citep{wang2020kinetic, su2023kinetic}. The resulting form of the collision term is given by

\begin{equation}
   \varOmega = J_{\rm S} + J_{\mathrm{e}}, 
\label{2.6}
\end{equation}
where the Shakhov model-like part $J_{\rm S}$ is 
\begin{equation}
   J_{\mathrm{S}} = \frac{f^{\mathrm{eq}}-f}{\tau} + \frac{f^\mathrm{eq}}{\tau}\frac{2m(1-Pr)\boldsymbol{q^{\rm K}}\cdot\boldsymbol{C}}{5n(k_{\rm B}T)^{2}}\left(\frac{mC^{2}}{2k_{\rm B}T}-\frac{5}{2}\right),
\label{2.7}
\end{equation}
and the excess part $J_{e}$ is
\begin{equation}
\begin{split}
     J_{\mathrm{e}} = &-\rho b\chi f^\mathrm{eq}\left\{\boldsymbol{C} \cdot \left[\frac{2}{n}\frac{\partial n}{\partial \boldsymbol{x}}+\frac{1}{T}\frac{\partial T}{\partial \boldsymbol{x}}\left(\frac{3mC^{2}}{10k_{\rm B}T}-\frac{1}{2}\right)\right] \right.\\
     & \left.+\frac{2m}{5k_{\rm B}T}\boldsymbol{CC}:\frac{\partial}{\partial \boldsymbol{x}}\boldsymbol{u} -\left(1-\frac{mC^{2}}{5k_{\rm B}T}\right)\left(\frac{\partial}{\partial \boldsymbol{x}}\cdot\boldsymbol{u}\right) \right\} \\
     &-\rho b \boldsymbol{C}f^\mathrm{eq} \cdot \frac{\partial \chi}{\partial \boldsymbol{x}} \\
     &+\frac{\partial}{\partial \boldsymbol{x}} \cdot \left[f^\mathrm{eq}\frac{\varpi}{nk_{\rm B}T}\left(\frac{\partial}{\partial \boldsymbol{x}} \cdot \boldsymbol{u}\right)\boldsymbol{C}\left(\frac{mC^{2}}{2k_{\rm B}T}-\frac{3}{2}\right)\right].
\label{2.8}
\end{split}
\end{equation}

Here, $\varpi$ denotes the bulk viscosity, and $f^\mathrm{eq}$ is the local Maxwellian distribution function, given by
\begin{equation}
   f^\mathrm{eq} =  n\left(\frac{m}{2\pi k_{\rm B}T}\right)^{3/2}{\rm exp}\left(-\frac{mC^{2}}{2k_{\rm B}T}\right),
\label{2.9}
\end{equation}
where $n$ is the number density, $T$ is the temperature, $\boldsymbol{u}$ is the bulk velocity, and $\boldsymbol{q}^{\rm K}$ is the kinetic contribution to the heat flux. These macroscopic quantities are obtained from the velocity moments of the distribution function. The equation of state (EoS) derived from the EV-form model takes a generalized van der Waals form:
\begin{equation}
   p = nk_{\rm B}T(1+\rho b \chi)-an^{2},
\label{eq 11}
\end{equation}
where $b = 2{\pi}\sigma^{3}/3m$, $\chi$ is the pair correlation function and $a$ is the Vlasov parameter. Both $\chi$ and $a$ are treated as adjustable constants. This equation describes the complex dependence of pressure on number density and temperature, and serves as the basis for determining the critical parameters of the fluid. Specifically, the critical point is identified using the standard conditions:
\begin{equation}
   \left(\frac{\partial p}{\partial n}\right)_{T=T_{\mathrm{c}}, n=n_\mathrm{c}} = 0, \quad \left(\frac{\partial^{2} p}{\partial n^{2}}\right)_{T=T_{\mathrm{c}}, n=n_\mathrm{c}} = 0.
\label{eq 12}
\end{equation}

To accurately reproduce the thermodynamic properties of Lennard-Jones fluids, we adopt the approach of Benilov and Benilov \citep{benilov2018energy, benilov2019peculiar} and modify the van der Waals form EoS by introducing a generalized approach for determining the the Vlasov parameter $a$ and pair correlation function $\chi$. This method ensures consistency with key reference data such as the critical and triple points. The Vlasov parameter $a$ is determined by fitting a linear relation to empirical data on the non-ideal component of the per-molecule internal energy of Lennard-Jones fluids. It also characterizes the strength of the molecular interaction forces. Meanwhile, the pair correlation function $\chi$ is derived from an empirical function $\mathit{\mathit{\Phi}}$, which characterizes the non-ideal contribution to the fluid’s entropy. In principle, $\mathit{\mathit{\Phi}}$ should vanish at zero density, i.e., $\mathit{\mathit{\Phi}}(0) = 0$, and satisfy $\mathit{\mathit{\Phi}}^{'}(0) = 1$ within the EV framework \citep{benilov2018energy}. The formulation of $\mathit{\mathit{\Phi}}$ is introduced as:
\begin{equation}
   \mathit{\mathit{\Phi}} = \rho b+a_{1}(\rho b)^{2}+a_{2}(\rho b)^{3}+a_{3}(\rho b)^{4}+a_{4}(\rho b)^{5},
\label{eq 13}
\end{equation}
which is related to $\chi$ through
\begin{equation}
   \chi = \mathit{\mathit{\Phi}}^{'} =  \frac{{\rm d} \mathit{\mathit{\Phi}}}{{\rm d}(\rho b)}.
\end{equation}
This formulation provides sufficient flexibility to fit multiple thermodynamic constraints simultaneously. To ensure consistency with the critical point, the parameters $a_{1,2,3,4}$ are calibrated using the Eq. (\ref{eq 12}). For the triple point, the per-molecule Gibbs free energy $G$ is introduced:
\begin{equation}
   G = k_{\rm B}T[{\rm ln}(nT^{-3/2})+ \mathit{\mathit{\Phi}} + \frac{3}{2\pi}\rho b \mathit{\mathit{\Phi}}^{'}] - 2an,
\label{eq 14}
\end{equation}
which is equivalent to the chemical potential for a single-component fluid. The coexistence conditions at the triple point require
\begin{equation}
   p(n_{\mathrm{v}},T_{\mathrm{tr}}) = p(n_{\mathrm{l}}, T_{\mathrm{tr}}), \quad G(n_{\mathrm{v}},T_{\mathrm{tr}}) = G(n_{\mathrm{l}}, T_{\mathrm{tr}}),
\label{eq 15}
\end{equation}
where the liquid-phase number density at the triple point is given by
\begin{equation}
   n_{{\mathrm{ltr}}}\sigma^{3} = 1.
\label{eq 16}
\end{equation}
By simultaneously solving Eqs. (\ref{eq 12}), (\ref{eq 15}), and (\ref{eq 16}), the empirical coefficients in Eq. (\ref{eq 13}) can be uniquely determined.

This methodology provides a general framework that can be applied to all Lennard-Jones fluids. As an example, we apply it to Argon. The relevant thermodynamic parameters and the fitted value of the Vlasov term are listed in Table~\ref{tbl1}.

\begin{table}
\caption{The Vlasov term, critical and triple point parameters of Argon. All values are normalized by Avogadro constant $N_{\rm A}$}\label{tbl1}
\begin{tabular*}{\tblwidth}{@{}LLLLL@{}}
\toprule
  $aN_{\rm A}^{2} {\hspace{2pt}} ({\rm Jl} \!\cdot\! {\rm mol}^{-2})$ & $n_\mathrm{c}/N_{\rm A}{\hspace{2pt}} (\rm mol\!\cdot\! l^{-1})$ & $T_{\mathrm{c}}{\hspace{2pt}}(\rm K)$ & $n_{\mathrm{tr}}/N_{\rm A} {\hspace{2pt}} (\rm mol \!\cdot\! l^{-1})$ & $T_{\mathrm{tr}} {\hspace{2pt}} (\rm K)$  \\ 
\midrule
162.5 & 13.41 & 150.69 & 35.47 & 83.81 \\
\bottomrule
\end{tabular*}
\end{table}
The fitted values for the empirical coefficients in Eq.~\ref{eq 13} are:
\begin{equation}
\begin{split}
   &a_{1} = -0.4133, \quad a_{2} = 1.0506, \\
   &a_{3} = -0.6693, \quad a_{4} = 0.1556.
\label{eq 17}
\end{split}
\end{equation}
Substituting these values into Eq. (\ref{eq 11}) yields a calibrated equation of state that accurately captures both the critical and triple-point thermodynamic behavior of Argon.

Finally, the relaxation time $\tau = \mu/nk_{\rm B}T$, while the shear viscosity of dense gas can be obtained by
\begin{equation}
   \mu = \frac{\mu^{*}}{\chi}\left(1+\frac{2}{5}\rho b\chi\right)^{2}+\frac{3}{5}\varpi,
\label{eq 18}
\end{equation}
where $\mu^{*}$ is the shear viscosity at the reference temperature, and the bulk viscosity is $\varpi = \mu^{*}\chi(\rho b)^{2}$. The thermal conductivity $\kappa$ for dense gas can be calculated through
\begin{equation}
   \kappa = \frac{\kappa^{*}}{\chi}\left(1+\frac{3}{5}\rho b\chi\right)^{2}+c_{\mathrm{v}}\varpi,
\label{eq 19}
\end{equation}
where $\kappa$ is the thermal conductivity at the reference pressure, $c_{\mathrm{v}} = 3k_{\rm B}/2m$ is the specific heat capacity at constant volume. Finally, the $Pr$ can be obtained through
\begin{equation}
   Pr = \frac{2}{3}\frac{(1+\frac{2}{5}\rho b \chi)^{2}+\frac{3}{5}(\rho b \chi)^{2}}{(1+\frac{3}{5}\rho b \chi)^{2}+\frac{2}{5}(\rho b \chi)^{2}}.
\label{eq 20}
\end{equation}

\section{Results and Discussion}
\label{section 3}

\begin{figure}[!htbp]
    \includegraphics[width=\linewidth]{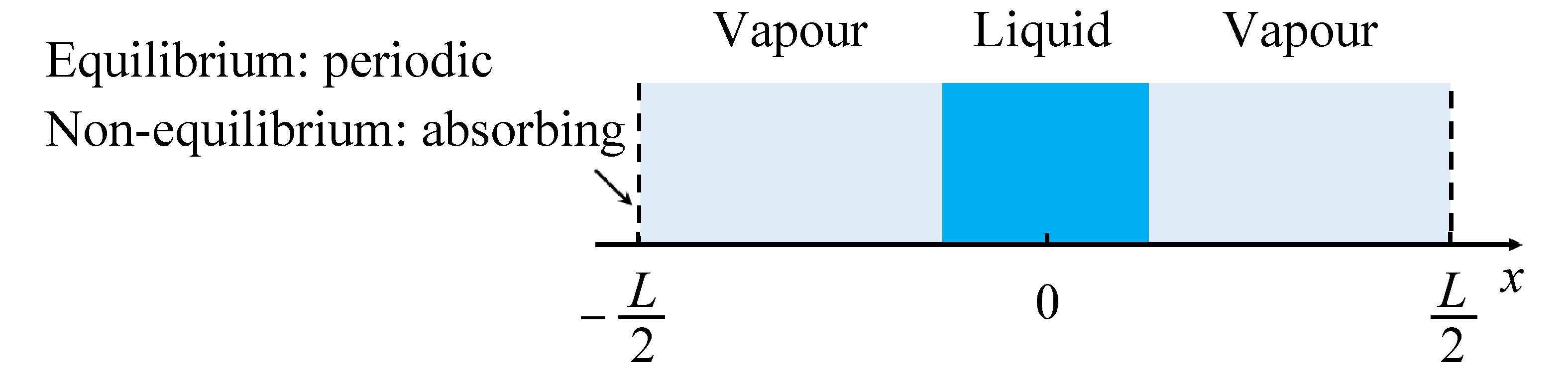}
    \caption{Schematic of the simulation setup, with boundary conditions indicated in the figure. Periodic boundary conditions are applied for equilibrium cases, while absorbing boundary conditions are used for non-equilibrium cases.}
    \label{fig1}
\end{figure}

The kinetic model is first validated by examining the equilibrium properties of a liquid–vapour system. To obtain the liquid–vapour coexistence curve, a liquid slab is initialized at the center of the simulation domain, surrounded by its vapour phase. The length of the computational domain is set as $L$. Periodic boundary conditions are applied on the left and right boundaries, and the temperature is maintained uniformly throughout the entire domain, as shown in Fig. \ref{fig1}. 

\begin{figure}[!htbp]
  \centering
    \includegraphics[width=\linewidth]{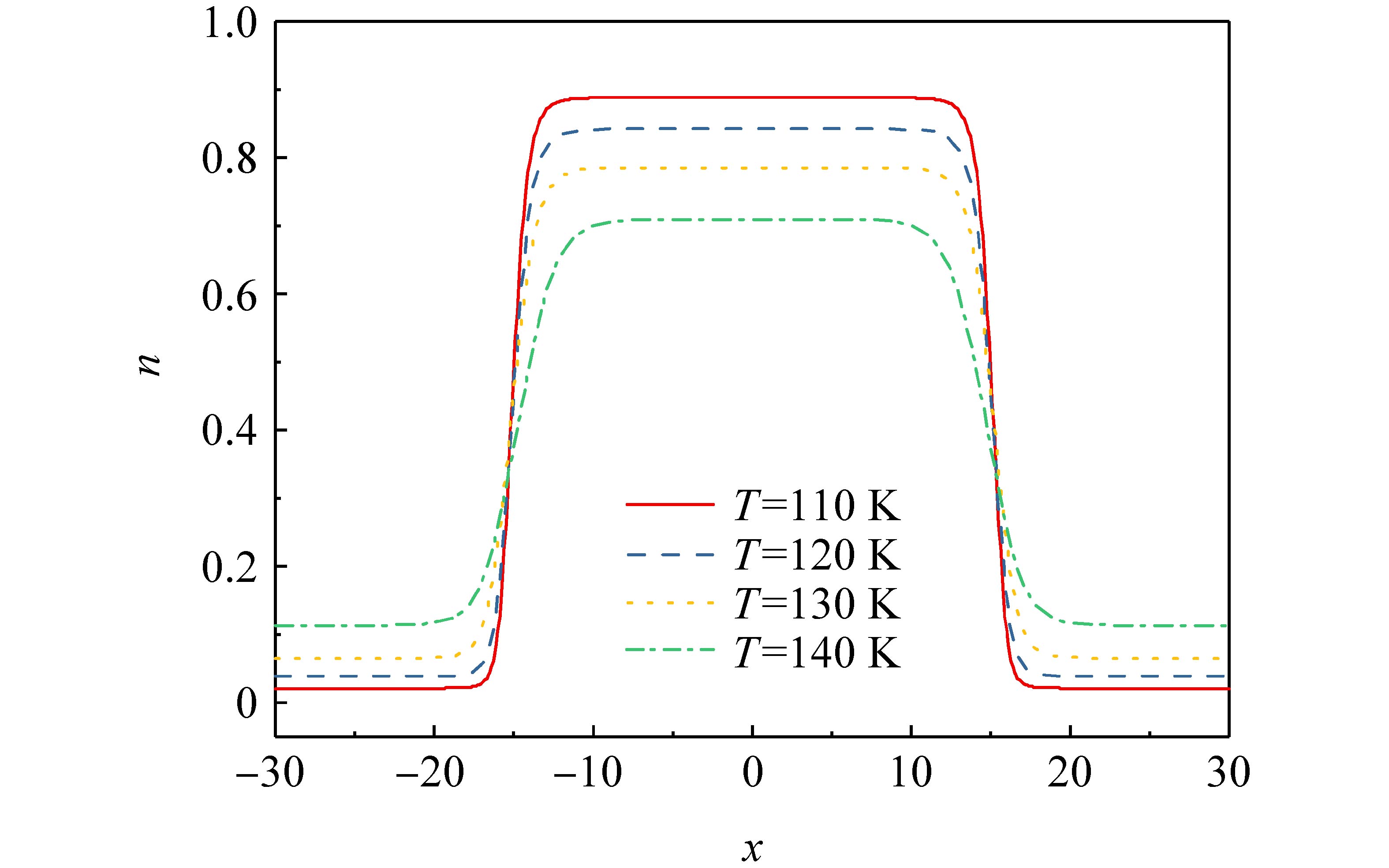}
    \caption{Equilibrium distributions of the number density at various temperatures.}
    \label{fig2}
\end{figure}

\begin{figure*}[!htbp]
  \centering
    \includegraphics[scale=1]{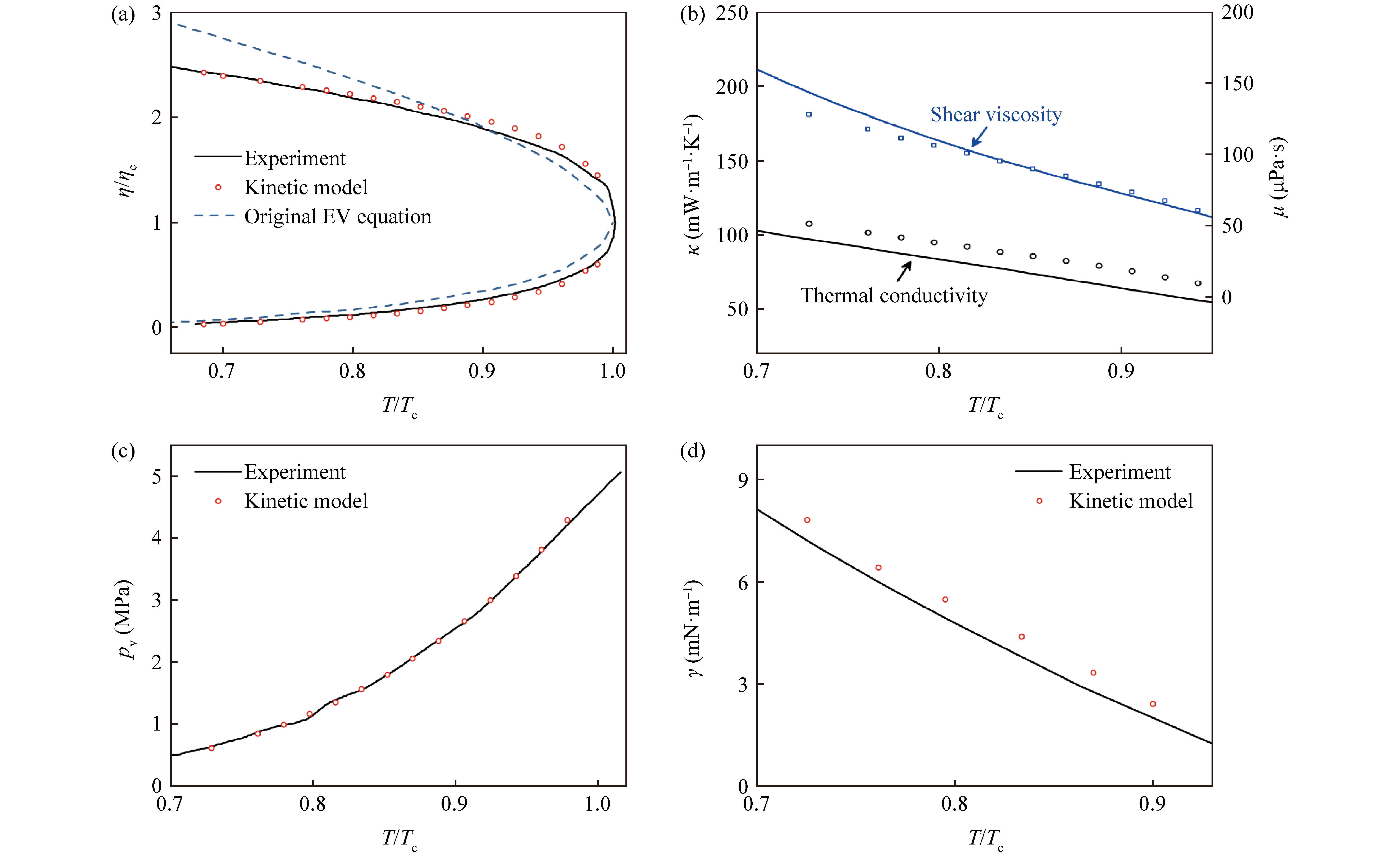}
    \caption{Comparison between the results obtained from kinetic model and experiment under different temperatures: (a) reduced number density; (b) shear viscosity and thermal conductivity; (c) vapour pressure; (d) surface tension coefficient. For better illustration, the liquid–vapour coexistence curve obtained from the original EV equation is shown in the figure 3(a). In figure 3(b), the results obtained from kinetic model are denoted by circles and squares, while the experimental data is denoted by solid lines.}
    \label{fig3}
\end{figure*}

To provide an intuitive representation of the simulated flow behavior, the equilibrium distributions of number density at various temperatures are shown in Fig. \ref{fig2} , where the distance is normalized by the molecular diameter, the number density is normalized by $1/\sigma^{3}$. It can be observed that the liquid number density decreases, while the vapour number density increases with rising temperature. The equilibrium bulk number densities are then compared with the results from the original EV equation and the experimental data for argon, as illustrated in Fig. \ref{fig3}(a). As seen in the figure, for the results obtained from the original EV equations, the number density of the liquid phase exhibits significant deviation from the experimental values especially for low temperatures. This discrepancy primarily originates from the overly simplistic repulsive term in the Sutherland potential used in the original formulation. After applying modifications to EoS, i.e., introducing a more accurate expression of the thermodynamic properties of Lennard-Jones fluids, excellent agreement with the experimental data has been achieved across a broad range of temperatures. In addition to the liquid-vapour coexistence curve, we also calculate the transport coefficients of the liquid bulk under saturation conditions and compare them with the experimental data \citep{younglove1986viscosity}. The shear viscosity and thermal conductivity are calculated using Eqs. (\ref{eq 18}) and (\ref{eq 19}) with the calibrated $\chi$. As shown in Fig. \ref{fig3}(b), the predicted shear viscosity agrees well with the experimental data across the investigated temperature range. For the thermal conductivity, some deviations are observed, with an average error of approximately $10\%$. We also examine the equilibrium vapour pressure, which is calculated from the model via Eq. (\ref{eq 11}). As illustrated in Fig. \ref{fig3}(c), the predicted vapour pressure is in good agreement with the experimental values over a broad range of temperatures, again confirming the validity of the kinetic model under equilibrium conditions. To further broaden the applicability of the model, we aim to evaluate the surface tension coefficient $\gamma$ by verifying the equilibrium properties of a single droplet. For this purpose, we extend the simulation to two dimensions and place a single droplet within the computational domain. The surface tension coefficient $\gamma$ is then obtained using the Young-Laplace equation
\begin{equation}
   p_{\mathrm{l}}-p_{\mathrm{v}} = \frac{\gamma}{R},
\label{eq 21}
\end{equation}
where $p_{\mathrm{l}}$ and $p_{\mathrm{v}}$ are the pressures in the liquid and vapour phases, respectively; and $R$ is the radius of the droplet. The computed surface tension coefficient is then compared with the experimental data, as shown in Fig. \ref{fig3}(d). The average relative error is approximately $9\%$. The close agreement between the simulation and experimental data in all four quantities, liquid-vapour coexistence curve, transport coefficient, vapour pressure, and surface tension, demonstrates that the proposed kinetic model can accurately capture the equilibrium behaviour of Lennard-Jones fluids in liquid-vapour systems. Slight overestimate of the surface tension coefficient may be attributed to the high-order curvature effect, which has been neglected in Eq. (\ref{eq 21}) \citep{lulli2022mesoscale}.

\begin{figure}[!htbp]
  \centering
    \includegraphics[scale=1]{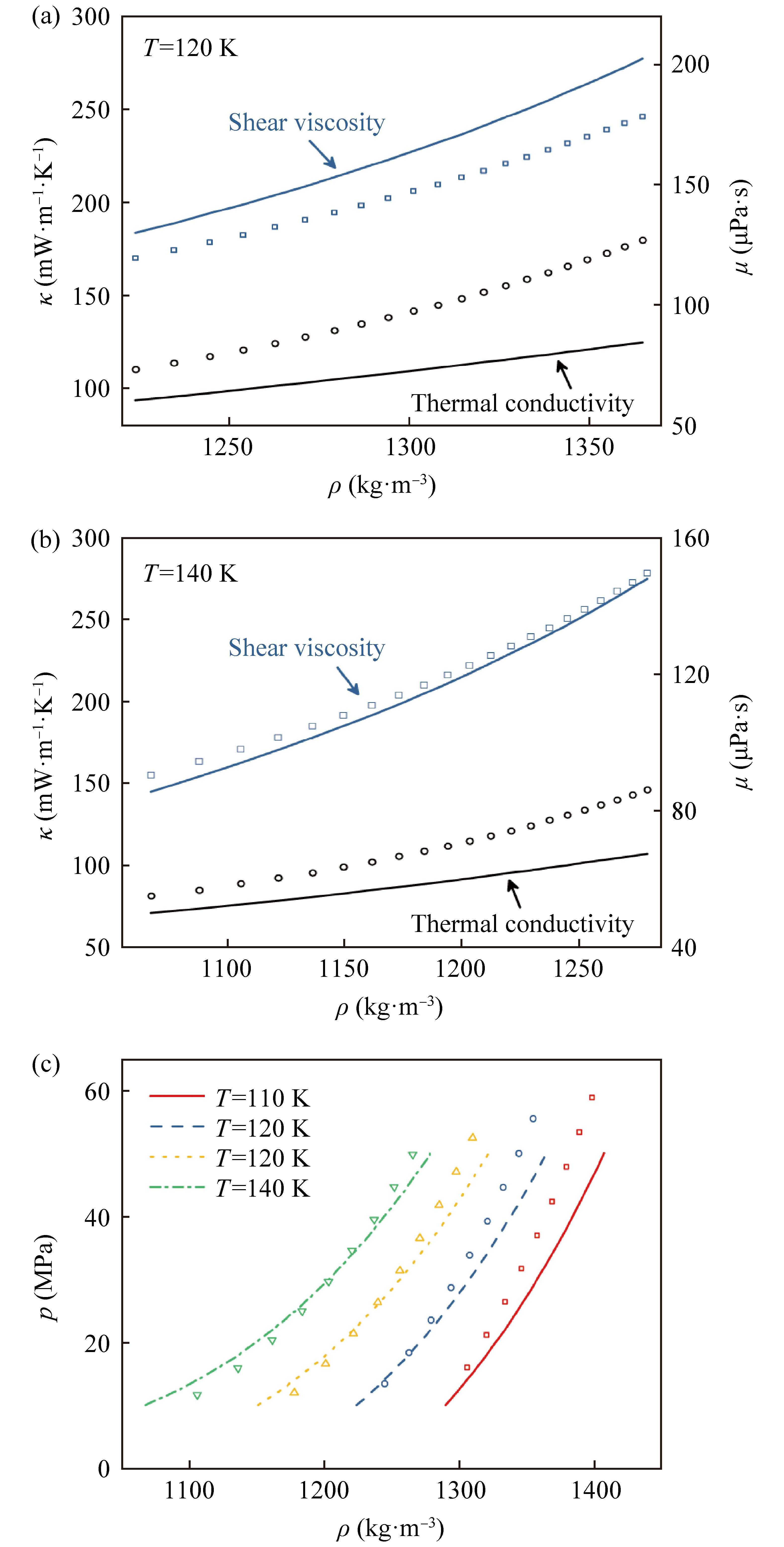}
    \caption{Comparison of transport coefficients (a, b) and pressure (c) at various temperatures and densities for the  liquid bulk. The kinetic model results are shown by the symbols, while the experimental data are represented by the lines.}
    \label{fig4}
\end{figure}

It is also important to assess whether the model can provide accurate predictions away from the saturation curve. Since unsaturated vapour can be treated as a dilute gas, we only present the results for the liquid bulk. Figure \ref{fig4} shows the transport coefficients and the pressure at various temperatures and densities in the liquid bulk. As shown in Fig. \ref{fig4} (a) and (b), the predicted shear viscosity agrees well with the experimental data at higher temperatures, while deviations increase at lower temperatures. The similar trend holds for thermal conductivity. For the pressure-density-temperature relations presented in Fig. \ref{fig4} (c), the predictions match the experimental data when the density is close to the saturation density and the temperature is near the critical temperature; larger deviations appear at higher densities and lower temperatures. These discrepancies are expected because the present equation of state is calibrated using only the critical and triple-point parameters and is primarily intended to reproduce saturation properties of the liquid-vapour system.

\begin{figure}[!h]
  \centering
    \includegraphics[width=\linewidth]{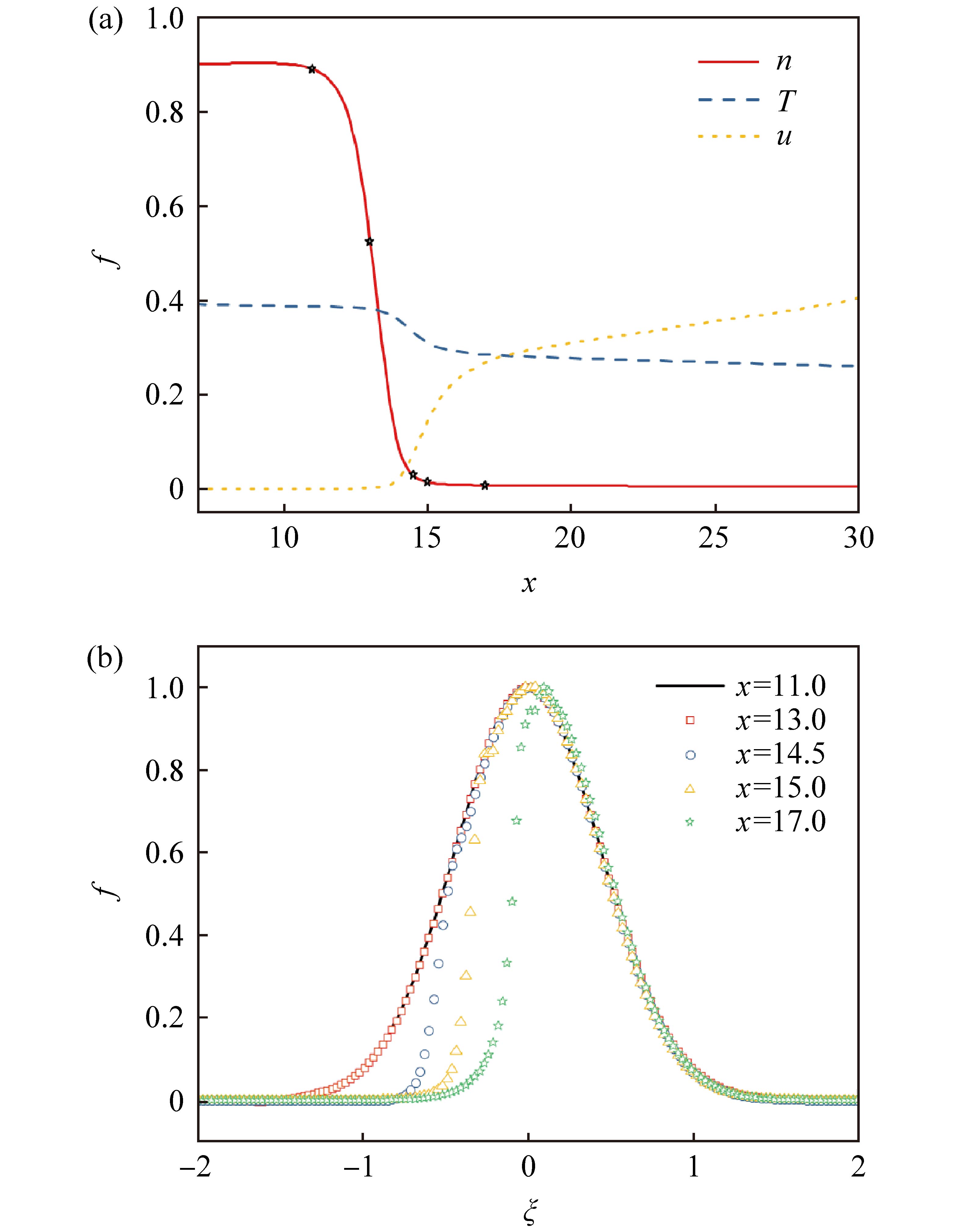}
    \caption{(a) Profiles of number density, temperature, and bulk velocity during evaporation into vacuum; (b) Velocity distribution functions at different positions: $x = 11, 13, 14.5, 15$ and $17$. These positions are also denoted by black hollow stars in the figure (a).}
    \label{fig5}
\end{figure}

After the model validation, we employ our kinetic model to simulate non-equilibrium evaporation process. The simulation setup is also given by Fig. \ref{fig1}. The absorbing walls are set on the boundaries, where the particles can only go outward and no particles are allowed to enter. The macroscopic profiles, including the number density, temperature, and bulk velocity, are shown in Fig. \ref{fig5}(a). In the figure, temperature is normalized by the room temperature, and the bulk velocity is normalized by the most probable molecular velocity. The difference to the equilibrium simulations is that the temperature decreases from the center to the boundary, while the number density has a slight increase due to the cooling effect. Since the particles are leaving through the boundaries, the velocity increases in the interface and vapour regions. 

In addition to the variation of macroscopic parameters, particular attention is given to the molecular velocity distribution function, as its characteristics can help explain why the classical HK relation becomes inaccurate under strong evaporation conditions. We select several positions for analysis: near the end of the liquid phase ($x = 11$), within the interface region ($x = 13$ ), at the early stage of the vapour phase ($x = 14.5$ and $x = 15$), and closer to the boundary in the vapour phase ($x = 17$). All the positions are also indicated by black hollow stars, from left to right, on the number density profile in Fig. \ref{fig5}(a). For better comparison, the distribution function $f$ is normalized by its maximum value at each position, and the molecular velocity $\xi$ is normalized by the molecular most probable velocity. As shown in Fig. \ref{fig5}(b), the velocity distribution function closely follows the Maxwellian distribution in the high density regions, including the liquid phase and interface region. However, as the position approaches the vapour region, deviations emerge in the negative velocity range due to insufficient molecular collisions in these low-density regions, while the positive velocity region still roughly follows the Maxwellian. Notably, the bulk velocity becomes increasingly pronounced as the position approaches the boundary, due to the continuous outflow of particles into the vacuum. The onset of deviation is clearly visible, providing evidence for the breakdown of the HK relation, which is caused by two key factors: first, the velocity distribution function deviates from the Maxwellian in the vapor region adjacent to the liquid-vapour interface; second, the influence of bulk motion becomes increasingly significant near the vapour side, leading to a noticeable shift of the distribution, as illustrated by the case at $x = 17$. Not only can these findings provide in-depth understanding of the flow characteristics of liquid–vapour systems under strong evaporation, but also offer valuable information for calibrating the HK relation in strong non-equilibrium evaporation.

\section{Conclusions}
\label{section4}

In this letter, we developed a kinetic model capable of accurately reproducing the fluid properties of Lennard-Jones fluids. Starting from the Enskog equation, we simplified the collision term using a Taylor expansion and modelled molecular interactions with the Sutherland potential. To capture the thermodynamic behavior of real fluids, we calibrated the equation of state by matching key parameters, namely, the critical temperature and density, and the triple-point temperature and density. The model was validated against the experimental and molecular simulation data for Argon, showing good agreement in predicting the liquid–vapour coexistence curve, transport coefficient, equilibrium vapour pressure, and surface tension coefficient. These results confirm the model is able to reproduce the essential thermodynamic properties of Lennard-Jones fluids. Furthermore, we applied the model to a non-equilibrium evaporative flow and analyzed the velocity distribution function at various spatial positions. The results reveal a clear deviation from the Maxwellian distribution in the negative velocity region of the beginning of vapour phase, highlighting inappropriateness of the equilibrium assumption underpinning the classical Hertz–Knudsen relation. The proposed approach can also be extended to other noble gases, since their intermolecular interactions are similar and the corresponding parameters can be determined in the same manner. Moreover, it has the potential to be generalized to more complex polyatomic fluids if the internal energy modes are not activated which is often the case for the phase transition problems.

\section*{Declaration of competing interest}
\label{section5}

The authors declare that they have no known competing financial interests or personal relationships that could have appeared to influence the work reported in this paper.


\bibliographystyle{unsrt}

\bibliography{cas-refs}

@article{dong2022high,
  title={High-yielding and stable desalination via photothermal membrane distillation with free-flow evaporation channel},
  author={Dong, Y. and Poredo{\v{s}}, P. and Ma, Q. and Wang, R.},
  journal={Desalination},
  volume={543},
  pages={116103},
  year={2022},
  publisher={Elsevier}
}

@article{vaartstra2020capillary,
  title={Capillary-fed, thin film evaporation devices},
  author={Vaartstra, G. and Zhang, L. and Lu, Z. and D{\'\i}az-Mar{\'\i}n, C. D. and Grossman, J. C. and Wang, E. N},
  journal={J. Appl. Phys.},
  volume={128},
  number={13},
  year={2020},
  publisher={AIP Publishing}
}

@article{hertz1882ueber,
  title={Ueber die Verdunstung der Fl{\"u}ssigkeiten, insbesondere des Quecksilbers, im luftleeren Raume},
  author={Hertz, H.},
  journal={Ann. Phys},
  volume={253},
  number={10},
  pages={177--193},
  year={1882},
  publisher={Wiley Online Library}
}

@article{knudsen1915maximale,
  title={Die maximale verdampfungsgeschwindigkeit des quecksilbers},
  author={Knudsen, M.},
  journal={Ann. Phys},
  volume={352},
  number={13},
  pages={697--708},
  year={1915},
  publisher={Wiley Online Library}
}

@article{labuntsov1979analysis,
  title={Analysis of intensive evaporation and condensation},
  author={Labuntsov, D. A. and Kryukov, A. P.},
  journal={Int. J. Heat Mass Transfer},
  volume={22},
  number={7},
  pages={989--1002},
  year={1979},
  publisher={Elsevier}
}

@article{meland2003evaporation,
  title={Evaporation and condensation {K}nudsen layers for nonunity condensation coefficient},
  author={Meland, R. and Ytrehus, T.},
  journal={Phys. Fluids},
  volume={15},
  number={5},
  pages={1348--1350},
  year={2003},
  publisher={American Institute of Physics}
}

@article{graur2021non,
  title={Non-equilibrium evaporation: 1{D} benchmark problem for single gas},
  author={Graur, I. A. and Gatapova, E. Y. and Wolf, M. and Batueva, M. A.},
  journal={Int. J. Heat Mass Transfer},
  volume={181},
  pages={121997},
  year={2021},
  publisher={Elsevier}
}

@article{john2019numerical,
  title={Numerical investigation of nanoporous evaporation using direct simulation {M}onte {C}arlo},
  author={John, B. and Enright, R. and Sprittles, J. E. and Gibelli, L. and Emerson, D. R. and Lockerby, D. A.},
  journal={Phys. Rev. Fluids},
  volume={4},
  number={11},
  pages={113401},
  year={2019},
  publisher={APS}
}

@article{john2021evaporation,
  title={Evaporation from arbitrary nanoporous membrane configurations: {A}n effective evaporation coefficient approach},
  author={John, B. and Gibelli, L. and Enright, R. and Sprittles, J. E. and Lockerby, D. A. and Emerson, D. R.},
  journal={Phys. Fluids},
  volume={33},
  number={3},
  year={2021},
  publisher={AIP Publishing}
}

@article{li2021theoretical,
  title={Theoretical and numerical study of nanoporous evaporation with receded liquid surface: effect of {K}nudsen number},
  author={Li, R. and Wang, J. and Xia, G.},
  journal={J. Fluid Mech.},
  volume={928},
  pages={A9},
  year={2021},
  publisher={Cambridge University Press}
}

@article{li2023effect,
  title={Effect of inter-pore interference on liquid evaporation rates from nanopores by direct simulation {M}onte {C}arlo},
  author={Li, R. and Yan, Z. and Xia, G.},
  journal={Phys. Fluids},
  volume={35},
  number={3},
  year={2023},
  publisher={AIP Publishing}
}

@article{persad2016expressions,
  title={Expressions for the evaporation and condensation coefficients in the {H}ertz-{K}nudsen relation},
  author={Persad, A. H. and Ward, C. A.},
  journal={Chemical reviews},
  volume={116},
  number={14},
  pages={7727--7767},
  year={2016},
  publisher={ACS Publications}
}

@book{enskog1922kinetische,
  title={Kinetische Theorie der W{\"a}rmeleitung: Reibung und Selbst-diffusion in Gewissen verdichteten gasen und fl{\"u}ssigkeiten},
  author={Enskog, D.},
  year={1922},
  publisher={Almqvist \& Wiksells boktryckeri-a.-b.}
}

@article{sobrino1967kinetic,
  title={On the kinetic theory of a van der {W}aals gas},
  author={Sobrino, L.},
  journal={Can. J. Phys.},
  volume={45},
  number={2},
  pages={363--385},
  year={1967},
  publisher={NRC Research Press Ottawa, Canada}
}

@article{grmela1971kinetic,
  title={Kinetic equation approach to phase transitions},
  author={Grmela, M.},
  journal={J. Stat. Phys.},
  volume={3},
  pages={347--364},
  year={1971},
  publisher={Springer}
}

@article{karkheck1981kinetic,
  title={Kinetic mean-field theories},
  author={Karkheck, J. and Stell, G.},
  journal={J. Chem. Phys.},
  volume={75},
  number={3},
  pages={1475--1487},
  year={1981},
  publisher={American Institute of Physics}
}

@article{frezzotti2005mean,
  title={Mean field kinetic theory description of evaporation of a fluid into vacuum},
  author={Frezzotti, A. and Gibelli, L. and Lorenzani, S.},
  journal={Phys. Fluids},
  volume={17},
  number={1},
  year={2005},
  publisher={AIP Publishing}
}

@article{barbante2015kinetic,
  title={A kinetic theory description of liquid menisci at the microscale},
  author={Barbante, P. F. and Frezzotti, A. and Gibelli, L.},
  journal={Kinet. Relat. Models},
  volume={8},
  number={2},
  pages={235--254},
  year={2015}
}

@article{frezzotti2017kinetic,
  title={Kinetic theory aspects of non-equilibrium liquid-vapor flows},
  author={Frezzotti, A. and Barbante, P.},
  journal={Mech. Eng. Rev.},
  volume={4},
  number={2},
  pages={16--00540},
  year={2017},
  publisher={The Japan Society of Mechanical Engineers}
}

@article{busuioc2020mean,
  title={Mean-field kinetic theory approach to Langmuir evaporation of polyatomic liquids},
  author={Busuioc, S. and Gibelli, L.},
  journal={Phys. Fluids},
  volume={32},
  number={9},
  year={2020},
  publisher={AIP Publishing}
}

@article{homes2025heat,
  title={Heat and mass transfer across the vapor-liquid interface: {A} comparison of molecular dynamics and the {E}nskog-{V}lasov kinetic model},
  author={Homes, S. and Frezzotti, A. and Nitzke, I. and Struchtrup, H. and Vrabec, J.},
  journal={Int. J. Heat Mass Transfer},
  volume={242},
  pages={126828},
  year={2025},
  publisher={Elsevier}
}

@article{wang2020kinetic,
  title={The kinetic {S}hakhov-{E}nskog model for non-equilibrium flow of dense gases},
  author={Wang, P. and Wu, L. and Ho, M. T. and Li, J. and Li, Z. and Zhang, Y.},
  journal={J. Fluid Mech.},
  volume={883},
  pages={A48},
  year={2020},
  publisher={Cambridge University Press}
}

@article{su2023kinetic,
  title={Kinetic modeling of nonequilibrium flow of hard-sphere dense gases},
  author={Su, W. and Gibelli, L. and Li, J. and Borg, M. K. and Zhang, Y.},
  journal={Phys. Rev. Fluids},
  volume={8},
  number={1},
  pages={013401},
  year={2023},
  publisher={APS}
}

@article{shan2023molecular,
  title={Molecular kinetic modelling of non-equilibrium transport of confined van der {W}aals fluids},
  author={Shan, B. and Su, W. and Gibelli, L. and Zhang, Y.},
  journal={J. Fluid Mech.},
  volume={976},
  pages={A7},
  year={2023},
  publisher={Cambridge University Press}
}

@article{li2024molecular,
  title={Molecular kinetic modelling of non-equilibrium evaporative flows},
  author={Li, S. and Su, W. and Shan, B. and Li, Z. and Gibelli, L. and Zhang, Y.},
  journal={J. Fluid Mech.},
  volume={994},
  pages={A16},
  year={2024},
  publisher={Cambridge University Press}
}

@article{benilov2018energy,
  title={Energy conservation and {H}-theorem for the {E}nskog-{V}lasov equation},
  author={Benilov, E. S. and Benilov, M. S.},
  journal={Phys. Rev. E},
  volume={97},
  number={6},
  pages={062115},
  year={2018},
  publisher={APS}
}

@article{benilov2019peculiar,
  title={Peculiar property of noble gases and its explanation through the {E}nskog-{V}lasov model},
  author={Benilov, E. S. and Benilov, M. S.},
  journal={Phys. Rev. E},
  volume={99},
  number={1},
  pages={012144},
  year={2019},
  publisher={APS}
}

@article{chung1988generalized,
  title={Generalized multiparameter correlation for nonpolar and polar fluid transport properties},
  author={Chung, T. H. and Ajlan, M. and Lee, L. L. and Starling, K. E.},
  journal={Ind. Eng. Chem. Res.},
  volume={27},
  number={4},
  pages={671--679},
  year={1988},
  publisher={ACS Publications}
}

@article{shan2025molecular,
  title={Molecular kinetic modelling of nanoscale confined flows},
  author={Shan, B. and Torrilhon, M. and Guo, Z. and Zhang, Y.},
  journal={J. Fluid Mech.},
  volume={1012},
  pages={A20},
  year={2025},
  publisher={Cambridge University Press}
}

@article{lulli2022mesoscale,
  title={Mesoscale perspective on the {T}olman length},
  author={Lulli, M. and Biferale, L. and Falcucci, G. and Sbragaglia, M. and Shan, X.},
  journal={Phys. Rev. E},
  volume={105},
  number={1},
  pages={015301},
  year={2022},
  publisher={APS}
}

@article{chen2024evaporation,
  title={Evaporation mechanisms during droplet levitation and coalescence based on molecular dynamics},
  author={Chen, F. and Gang, T. and Chen, L.},
  journal={Theor. Appl. Mech. Lett.},
  volume={14},
  number={6},
  pages={100533},
  year={2024},
  publisher={Elsevier}
}

@article{wang2025molecular,
  title={Molecular understanding of phase behavior of hydrocarbon mixtures in nanopores and their influence on recovery dynamics},
  author={Wang, J. and Wang, Y. and Li, B. and Wang, Q. and Meng, S. and Chen, R. and Wu, H. and Wang, F.},
  journal={Theor. Appl. Mech. Lett.},
  volume={15},
  number={4},
  pages={100589},
  year={2025},
  publisher={Elsevier}
}

@article{frezzotti2007numerical,
  title={A numerical investigation of the steady evaporation of a polyatomic gas},
  author={Frezzotti, A.},
  journal={Eur. J. Mech. B Fluids},
  volume={26},
  number={1},
  pages={93--104},
  year={2007},
  publisher={Elsevier}
}

@article{ytrehus1997molecular,
  title={Molecular-flow effects in evaporation and condensation at interfaces},
  author={Ytrehus, T.},
  journal={Multiph. Sci.},
  volume={9},
  number={3},
  year={1997},
  publisher={Begel House Inc.}
}

@article{vaartstra2022revisiting,
  title={Revisiting the Schrage equation for kinetically limited evaporation and condensation},
  author={Vaartstra, G. and Lu, Z. and Lienhard, J. H. and Wang, E. N.},
  journal={J. Heat Transf.},
  volume={144},
  number={8},
  pages={080802},
  year={2022},
  publisher={American Society of Mechanical Engineers}
}

@book{schrage1953theoretical,
  title={A theoretical study of interphase mass transfer},
  author={Schrage, R. W.},
  year={1953},
  publisher={Columbia University Press}
}

@article{younglove1986viscosity,
  title={The viscosity and thermal conductivity coefficients of gaseous and liquid argon},
  author={Younglove, B.A. and Hanley, H.J.M.},
  journal={J. Phys. Chem. Ref. Data},
  volume={15},
  number={4},
  pages={1323--1337},
  year={1986},
  publisher={American Institute of Physics for the National Institute of Standards and~…}
}

@article{ytrehus1996kinetic,
  title={Kinetic theory approach to interphase processes},
  author={Ytrehus, T. and {\O}stmo, S.},
  journal={Int. J. Multiphase Flow},
  volume={22},
  number={1},
  pages={133--155},
  year={1996},
  publisher={Elsevier}
}

@article{oskouei2025nonlinear,
  title={Nonlinear mass and heat transfer across liquid-vapor interfaces},
  author={Oskouei, P. F. and Struchtrup, H.},
  journal={Phys. Rev. E},
  volume={112},
  number={2},
  pages={025501},
  year={2025},
  publisher={APS}
}

@article{hansen1969phase,
  title={Phase transitions of the {L}ennard-{J}ones system},
  author={Hansen, J. and Verlet, L.},
  journal={Phys. Rev.},
  volume={184},
  number={1},
  pages={151},
  year={1969},
  publisher={APS}
}

@article{tee1966molecular,
  title={Molecular parameters for normal fluids. {L}ennard-{J}ones 12-6 Potential},
  author={Tee, L. S. and Gotoh, S. and Stewart, W. E.},
  journal={Ind. Eng. Chem. Fund.},
  volume={5},
  number={3},
  pages={356--363},
  year={1966},
  publisher={ACS Publications}
}

\end{document}